\documentstyle[aps,preprint]{revtex}

\begin{document}

\title{Weak--coupling calculation of the gap structure \\
of doped $n$--leg Hubbard ladders \\}

\author{ N.\ Bulut and D.J.\ Scalapino }

\address{
Department of Physics, University of California \\
Santa Barbara, CA 93106--9530} 

\maketitle 
\begin{abstract}
In weak coupling, 
the spin gap in doped, even, $n$--leg periodic Hubbard ladders reflects 
the energy to break a pair into separate quasiparticles.
Here we investigate the structure of the gap within a spin--fluctuation 
exchange approximation.
We also calculate the amplitude for removing a singlet pair from two lattice
sites separated by a distance $(\ell_x,\ell_y)$, 
which describes the internal structure of a pair.
\end{abstract}

\pacs{PACS Numbers: 71.10.Fd, 71.10.Li, and 74.20.-z}

The half--filled two--leg Hubbard ladder has been found to have 
a spin--gap \cite{White}.  
Furthermore, near half--filling, when the interchain hopping 
is less than twice the intrachain hopping, holes doped into the 
system form singlet pairs and a reduced spin gap 
remains \cite{Noack,Balents}.
However, as first noted by Sigrist {\it et al.} \cite{Sigrist}, 
for the $t-J$ two--leg ladder,
the spin gap in the doped system differs  from the spin gap of the 
half--filled (undoped)
system and reflects the energy to separate a pair of holes into 
two quasiparticles.
This is most easily understood in the strong 
coupling limit when the rung exchange $J_\perp$ is
large compared to the interchain exchange coupling and the hopping $t$.
In this case, the ground state of the undoped ladder consists 
of singlet spin states on each rung and the spin gap is 
associated with exciting one of the rung singlets to a triplet.
In the doped ground state there are pairs of holes occupying 
various rungs. 
In this case, besides the triplet excitation of two spins on a rung, 
a singlet pair can be broken into two quasiparticles 
which are in a triplet spin state.
The separate delocalization reduction of the kinetic energy of the 
two quasiparticles makes it the lowest energy triplet state.
Thus the spin gap is set by the sum of the minimum quasiparticle 
energies.
Numerical calculations show that this remains true outside the 
strong coupling limit for the two--leg $t$--$J$ and Hubbard ladders.
A possible weak coupling view of the spin gap then is that it 
represents $2|\Delta_{\bf p}|$ with ${\bf p}=(p_F,0)$
or $(p_F,\pi)$.
Here $p_y=0$ or $\pi$ corresponds to the bonding or antibonding 
states of the 2--leg ladder.
If $\Delta_{\bf p}$ had the simple $d_{x^2-y^2}$ form 
$(\Delta_0/2)(\cos p_x - \cos p_y)$,
then $|\Delta_{\bf p}|$ would be non--vanishing for both 
$(p_F,0)$ and $(p_F,\pi)$.

As the number of legs increases, the size of the spin gap in the 
half--filled insulating case decreases.
Thus it could turn out that, the lowest triplet excitation of the doped 
Hubbard or $t$--$J$ models is similar to the undoped system and 
simply corresponds to a triplet excitation of the spin background 
rather than to breaking a pair.  
However, for the 4--leg ladder $p_y=0$, $\pm\pi/2$, $\pi$
and near half--filling, one can excite quasiparticles at 
$(\pm\pi/2,\pm\pi/2)$ where the $d_{x^2-y^2}$ gap vanishes.
This would suggest that, at least in weak coupling, near half--filling
the spin gap of the doped 4--leg Hubbard model vanishes even though 
the insulating 4--leg Hubbard system exhibits a spin gap.
Proceeding with the same type of argument for the 6--leg ladder 
where one can probe $p_y=0$, $\pm\pi/3$, $\pm 2\pi/3$, $\pi$,
the energy associated with creating two quasiparticles would be 
finite and thus there would be a finite spin gap in the doped
6--leg ladder set either by the triplet excitation of the spin 
background or the two quasiparticle triplet excitation. 

Here we investigate the structure of the gap in periodic $n$--leg ladders
within a weak coupling spin--fluctuation exchange approximation previously
used for the two dimensional Hubbard model.
We are also interested in calculating the internal structure of the 
pair wave function because it can be directly compared 
with the results of numerical Monte Carlo and 
density matrix renormalization group (DMRG) calculations. 

The Hubbard Hamiltonian is given by 
\begin{eqnarray}
H=-t\sum_{i,j,\sigma} (c_{i+1,j\sigma}^{\dagger} c_{i,j\sigma} +
c_{i,j\sigma}^{\dagger} c_{i+1,j\sigma} )
 - t_{\perp}\sum_{i,j,\sigma} (c_{i,j+1\sigma}^{\dagger} c_{i,j\sigma} + 
c_{i,j\sigma}^{\dagger} c_{i,j+1\sigma} )
 + U\sum_{i,j} n_{i,j\uparrow}n_{i,j\downarrow}.
\label{Hubbard}
\end{eqnarray}
Here $c_{i,j\sigma}^{\dagger}$ creates an electron of spin $\sigma$ on the 
$(i,j)$ lattice site and 
$n_{i,j\uparrow}= c_{i,j\uparrow}^{\dagger} c_{i,j\uparrow}$.
The index $j=1,...,n$ denotes a given leg of the ladder,
while $i$ denotes the position along a leg
which we took as 32 sites long. 
In addition,
periodic boundary conditions were used.
The hopping along a leg is $t$, the hopping between legs is $t_{\perp}$
and the onsite Coulomb interaction is $U$.
We used a spin--fluctuation interaction 
\begin{equation}
V({\bf q},i\omega_m)= {3\over 2}U^2\chi({\bf q},i\omega_m)
\label{V}
\end{equation}
with 
\begin{equation}
\chi({\bf q},i\omega_m)={ \chi_0({\bf q},i\omega_m) \over
1-U\chi_0({\bf q},i\omega_m) }
\label{chi}
\end{equation}
and 
\begin{equation}
\chi_0({\bf q},i\omega_m)= {1\over N} \sum_{{\bf p}}
{ f(\varepsilon_{{\bf p}+{\bf q}}) - f(\varepsilon_{{\bf p}}) \over
  i\omega_m - (\varepsilon_{{\bf p}+{\bf q}} - \varepsilon_{\bf p}) }.
\label{chi0}
\end{equation}
Here $\varepsilon_{\bf p} = -2t \cos{p_x} - 
2t_{\perp}\cos{p_y}-\mu$, $\omega_m=2m\pi T$,
$N=n\times 32$ is the number of lattice sites,
$\mu$ is the chemical potential,
and $f$ is the usual fermi factor.
The effective Coulomb interaction $U$ was adjusted to give 
strong antiferromagnetic spin fluctuations in
$\chi({\bf q},i\omega_m)$ and the interaction 
$V({\bf q},i\omega_m)$.

In order to model the relative internal structure 
of the pairing fluctuations in an $n$--leg ladder we have studied
the eigen solutions of the Bethe--Salpeter equation
\begin{equation}
-{T \over N}\sum_{p'} V(p-p') G_{\uparrow}(p') G_{\downarrow}(-p')
\phi(p') = \lambda \phi(p).
\label{BS}
\end{equation}
Here $p=({\bf p},i\omega_{n})$ and we have chosen $T$ such that 
the leading eigenvalue is less than unity.
In the following we have used the bare one--electron Green's functions
$G^{-1}({\bf p},i\omega_n)=i\omega_n-\varepsilon_{\bf p}$
in Eq.~(\ref{BS}).
While more elaborate conserving calculations are clearly 
important in estimating $T_c$, here we are interested in 
developing a picture of the internal structure of a pair.
We want guidance for Monte Carlo and
DMRG calculations and insight into the evolution of the 
spin gap behavior of doped $n$--leg ladders.

We first consider the 2--leg case.  The bonding $(p_y=0)$ and antibonding 
$(p_y=\pi)$ bands $\varepsilon_{\bf p} = -2t \cos{p_x} 
\pm 2t_{\perp}-\mu $ are plotted in Fig. 1 for $t_{\perp}=0.5t$ and a filling 
$\langle n\rangle = 1.0$.
The RPA spin susceptibility 
$\chi({\bf q},0)$, given by Eq.~(\ref{chi}) with $U=1.5t$ is plotted in
Fig. 2.
The strong antiferromagnetic fluctuations at ${\bf q}=(\pi,\pi)$ for a 
temperature $T=0.1t$ are clearly seen in Fig. 2.
The eigenfunction $\phi({\bf p},i\omega_n)$ 
with the biggest eigenvalue at this temperature is plotted in Figure 3
as a function of $p_x$ for 
$p_y=0$ and $\pi$, and $\omega_n=\pi T$ \cite{norm}.
As expected, $\phi({\bf p},i\pi T)$ peaks near the fermi 
surface and has opposite signs on the bonding and antibonding 
fermi points reflecting the $d_{x^2-y^2}$--like behavior 
of the bound state of two holes doped into a half--filled ladder.
As discussed, since only the wave vectors $(p_f,0)$ and 
$(p_f,\pm \pi)$ are allowed, the quasiparticle spectrum is gapped.
Thus for the 2--leg ladder a finite spin gap remains,
consistent with the DMRG \cite{Noack} and renormalization 
group \cite{Balents} calculations for the doped 2--leg Hubbard ladder.

The amplitude describing the internal structure of a singlet pair is 
\begin{eqnarray}
A(\ell_x,\ell_y)= \langle \Psi_0^{N-2}| 
( c_{i+\ell_x,j+\ell_y\uparrow} c_{i,j\downarrow} 
- c_{i,j\downarrow} c_{i+\ell_x,j+\ell_y\uparrow}) | \Psi_0^N\rangle.
\label{A}
\end{eqnarray}
Here $A(\ell_x,\ell_y)$ is the amplitude for removing a singlet pair from 
two sites separated by $(\ell_x,\ell_y)$.  
Using the solution of the Bethe--Salpeter equation, we have 
\begin{equation}
A(\ell_x,\ell_y)= {T \over N} \sum_{{\bf p},i\omega_n}
{\phi({\bf p},i\omega_n) \over \omega_n^2 + \varepsilon_{\bf p}^2}
e^{i {\bf p}\cdot {\ell}}
\label{A2}
\end{equation}
which is shown in Fig. 4.  
The relative $d_{x^2-y^2}$ like structure in which the amplitude at 
$(\ell_x=\pm 1,\ell_y=0)$ is out of phase with that at 
$(\ell_x=0,\ell_y =\pm 1)$ is clearly seen. 

We have carried out similar calculations for $n=4$ and 6 leg ladders.
Figure 5(a) shows the four band $p_x$ dispersion for the 
$n=4$ leg ladder and below it in Figure 5(b) the leading Bethe--Salpeter
eigenfunction $\phi({\bf p},i\omega_n)$
for $\omega_n=\pi T$.
In this case the eigenfunction has peaks with opposite signs at 
the $p_y=0$ and $p_y=\pi$ fermi surface points and vanishes 
at $p_y= \pi/2$.
This latter point corresponds to $(p_x=\pi/2,p_y=\pi/2)$
and represents a node in the $d_{x^2-y^2}$--like eigenfunction.
The amplitude $A(\ell_x,\ell_y)$ is plotted in Fig. 6 and one sees the 
characteristic $d_{x^2-y^2}$-like pattern with the 45 degree nodes.
Similar results for the 6--leg ladder are shown in Figures 7 and 8.
In this case, allowed values of $p_y$ are 0, $\pm \pi/3$,
$\pm 2\pi/3$, and $\pi$.
Again, large values of $\phi({\bf p},i\pi T)$ 
occur at the fermi surface for $p_y=0$ and $\pi$.
For $p_y=\pm \pi/3$ and $\pm 2\pi/3$, the gap is small but non--vanishing. 
The amplitude $A(\ell_x,\ell_y)$ for the 6--leg ladder is plotted in 
Fig. 8.  All of these results were for $\langle n\rangle =1.0$,
and we believe reflect the internal structure of the pair formed
when two holes are added to a half--filled $n$--leg ladder. 

Figure 9 shows the dispersion relations and Bethe--Salpeter 
amplitude $\phi({\bf p},i\pi T)$ for a 2--leg ladder with 
$\langle n\rangle = 0.85$.
Here the two peaks in the bonding and antibonding bands arise from 
umklapp processes.
Note that the relative sign difference remains. 
The relative internal pair amplitude 
$(\ell_x,\ell_y)$ for this case is shown in Fig. 10.

Within this approach, the Bethe--Salpeter eigenfunction provides a
measure of the quasiparticle gap and an $n$--leg ladder allows one to 
examine this gap for $p_y=0,\pi/n,2\pi/n,...,\pi$.
This is illustrated in Figure 11 for $n=6$.
Here the fermi surface for the $n\rightarrow \infty$ 
two--dimensional lattice with $t_{\perp}/t=0.5$ is shown as the dashed 
curve and $\phi(p_x,p_y,i\pi T)$ is shown versus $p_x$ for $p_y=0$,
$\pm \pi/3$, $\pm 2\pi/3$ and $\pm \pi$.
It is clear that within this approximation the spin gap of the doped
$n$--leg ladder is a reflection of the $d_{x^2-y^2}$--like
quasiparticle gap.
It will be interesting to see to what extent 
Monte Carlo and DMRG results 
support this picture.

%\vskip 0.2in
%\newpage
%\nonum

\acknowledgments

The authors gratefully acknowledge support from 
the National Science Foundation under Grant No. DMR92--25027.
The numerical computations reported in this 
paper were performed at the San Diego Supercomputer Center.

%\newpage

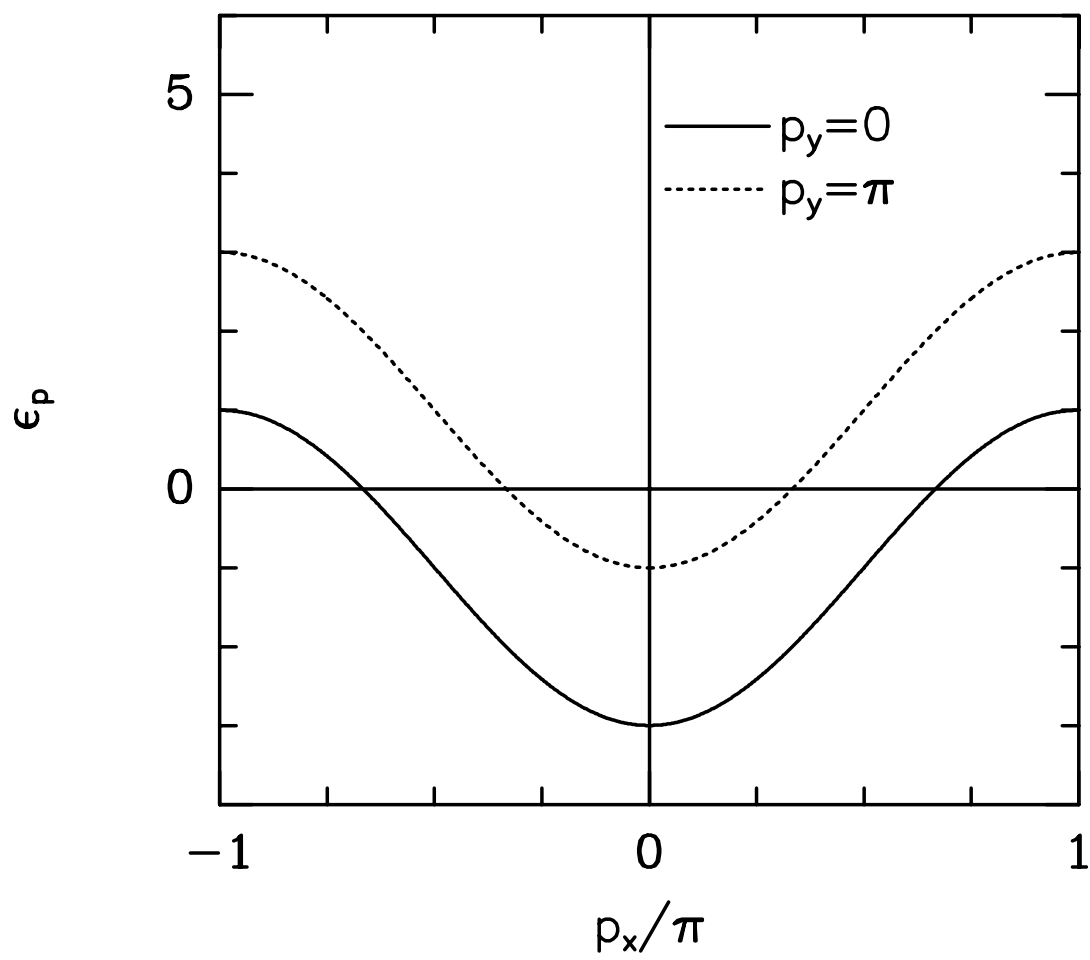
\begin{figure}
\caption{Quasiparticle dispersion $\varepsilon_{{\bf p}}$
versus $p_x$ for $p_y=0$ and $\pi$.
These results are for a two--leg ladder with 
$t_{\perp}=0.5t$ and $\langle n\rangle =1.0$.
}
\label{fig1}
\end{figure}
  
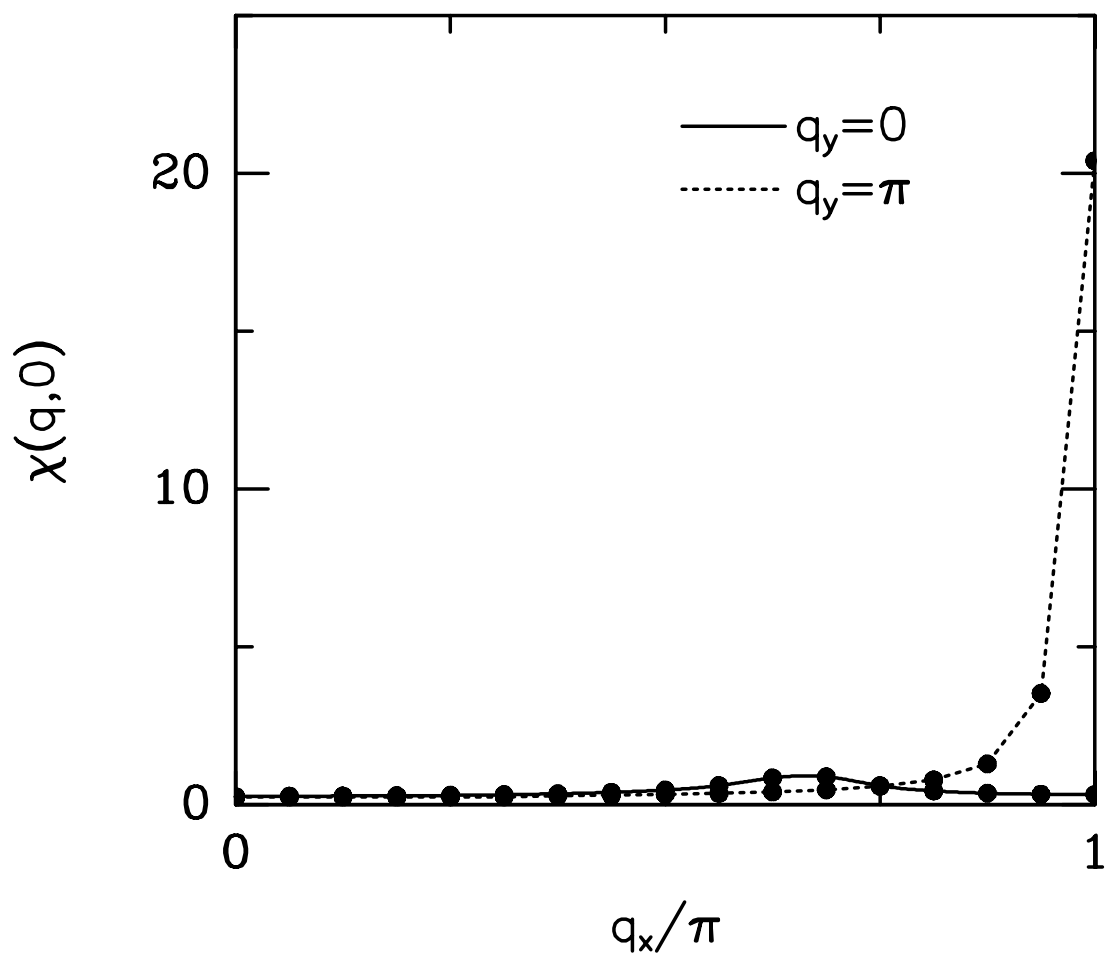
\begin{figure}
\caption{Magnetic susceptibility $\chi({\bf q},0)$ 
versus $q_x$ for $q_y=0$ and $\pi$.
These results are for a two--leg ladder with
$U=1.5t$, $t_{\perp}=0.5t$, $\langle n\rangle =1.0$
and $T=0.1t$.
}
\label{fig2}
\end{figure}
  
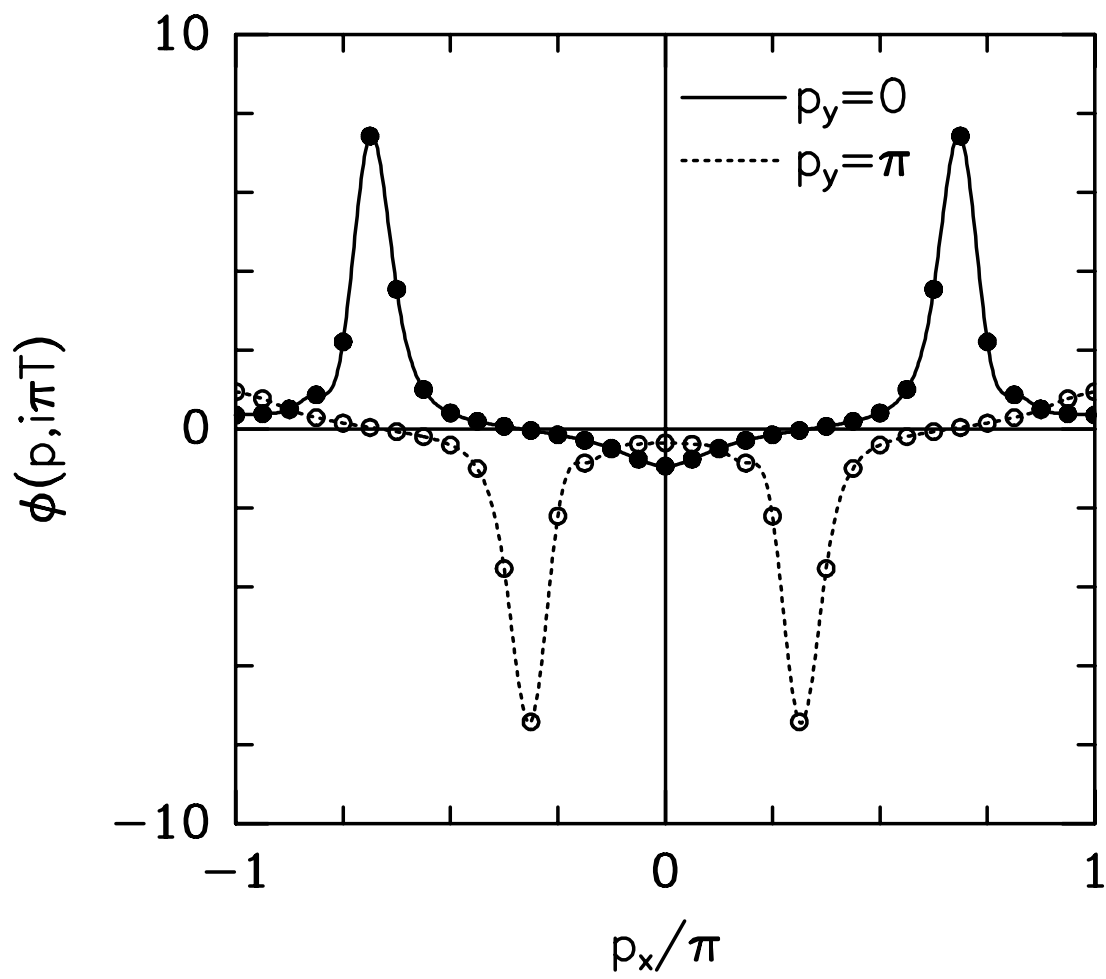
\begin{figure}
\caption{$\phi({\bf p},i\pi T)$ versus $p_x$ for 
$p_y=0$ and $\pi$.
These results are for the same parameters as in Fig. 2.
}
\label{fig3}
\end{figure}
  
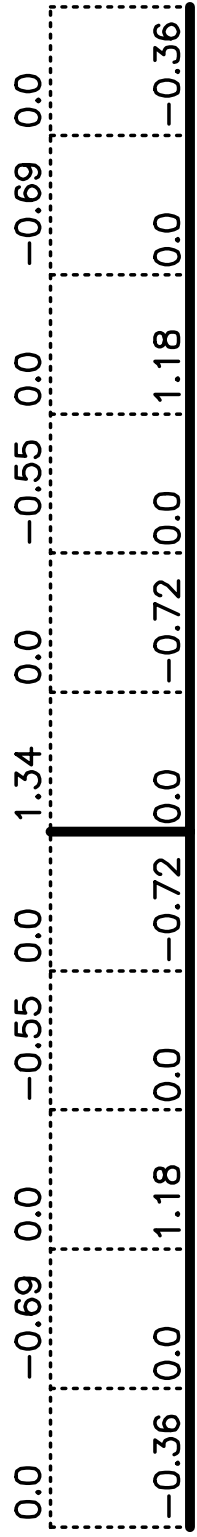
\begin{figure}
\caption{(a) Pair amplitude $A(\ell_x,\ell_y)$ versus 
$(\ell_x,\ell_y)$ for the same parameters as in Fig. 2.
In this figure the intersection of the thick solid lines 
corresponds to the origin of the $2\times 32$ lattice,
$(\ell_x,\ell_y)=(0,0)$.
}
\label{fig4}
\end{figure}
  
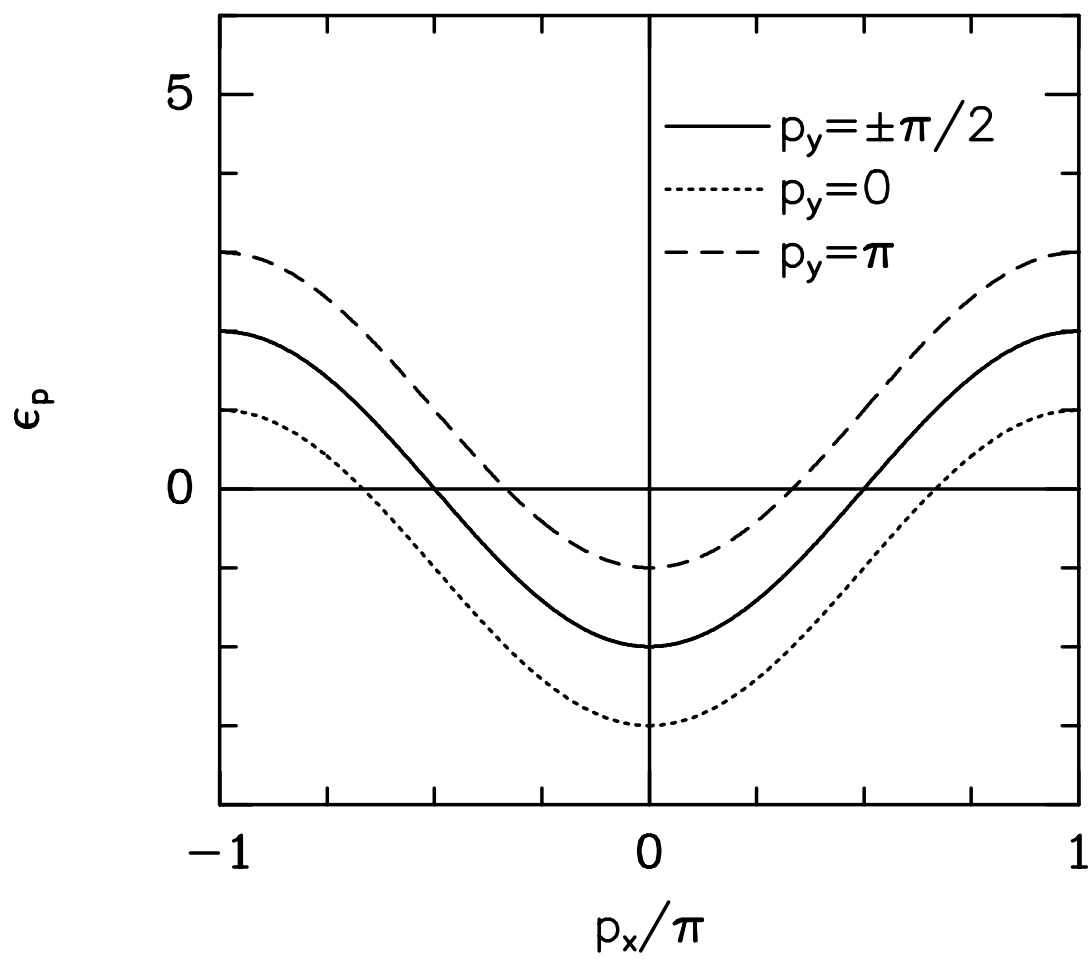
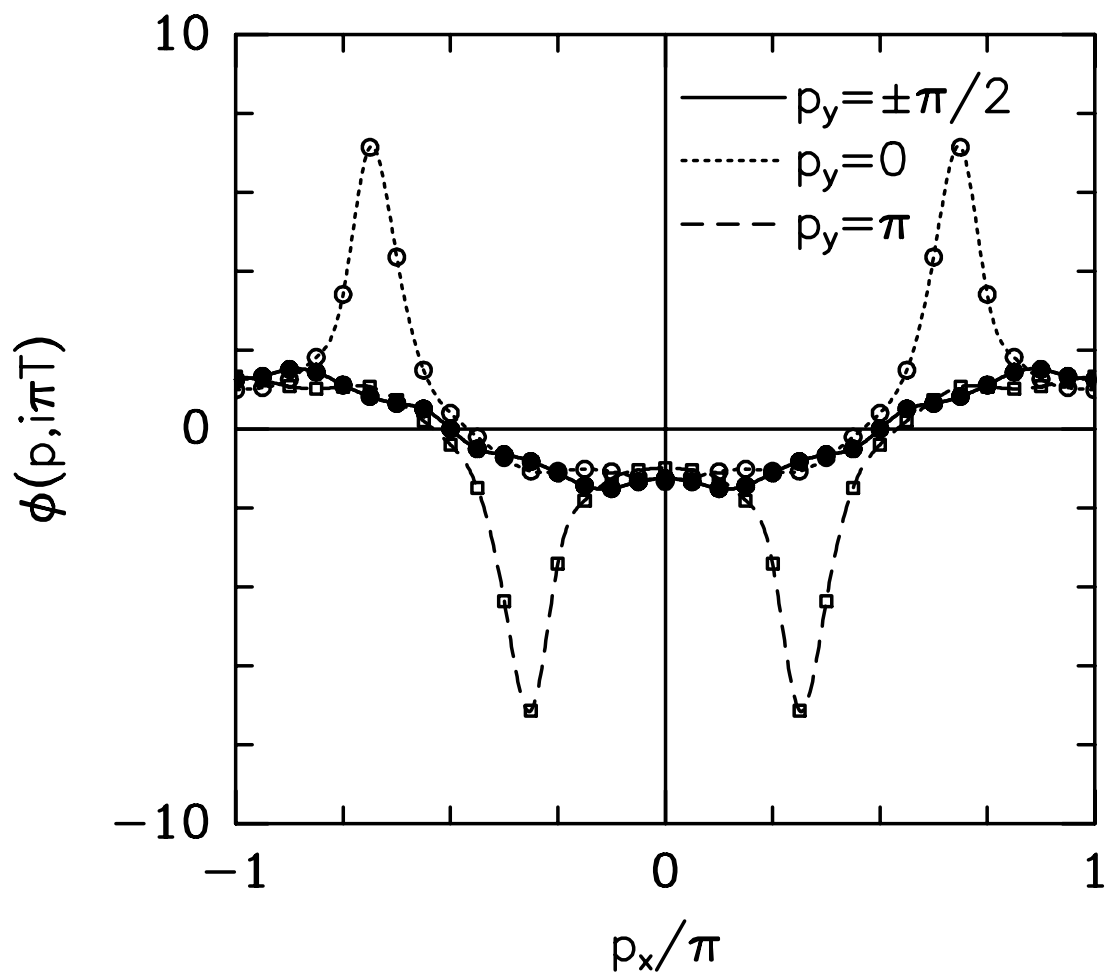
\begin{figure}
\caption{
(a) Quasiparticle dispersion $\varepsilon_{{\bf p}}$, and
(b) $\phi({\bf p},i\pi T)$ versus $p_x$
for different values of $p_y$.
These results are for a four--leg ladder with
$U=1.5t$, $t_{\perp}=0.5t$,
$\langle n\rangle = 1.0$, and $T=0.1t$.
}
\label{fig5}
\end{figure}
  
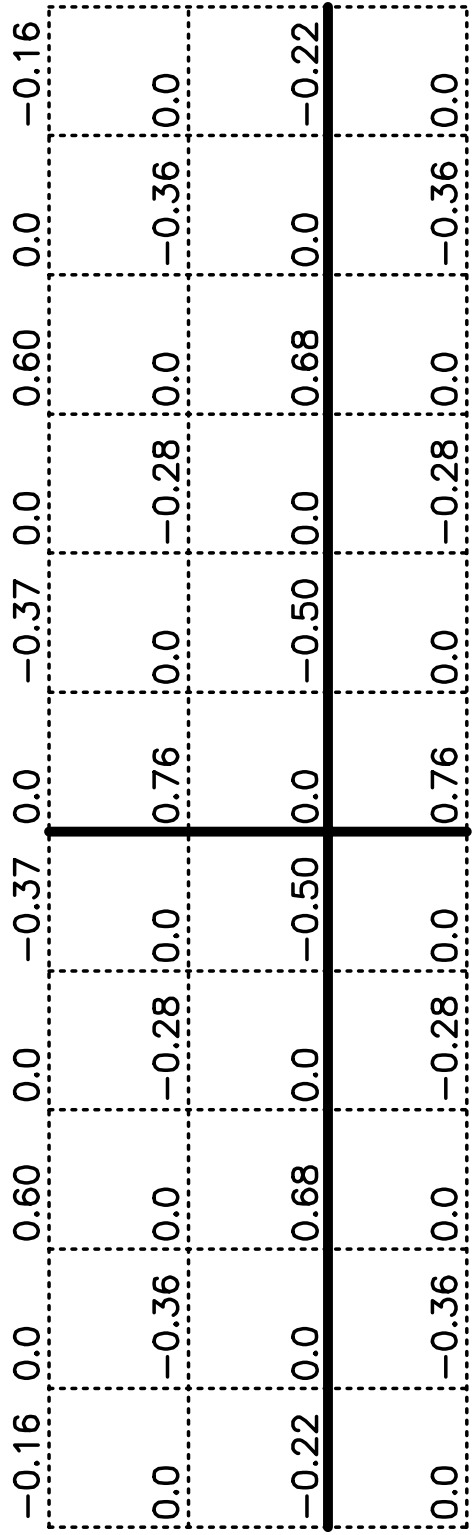
\begin{figure}
\caption{
Pair amplitude $A(\ell_x,\ell_y)$ versus $(\ell_x,\ell_y)$
for the same parameters as in Fig. 5.
}
\label{fig6}
\end{figure}
  
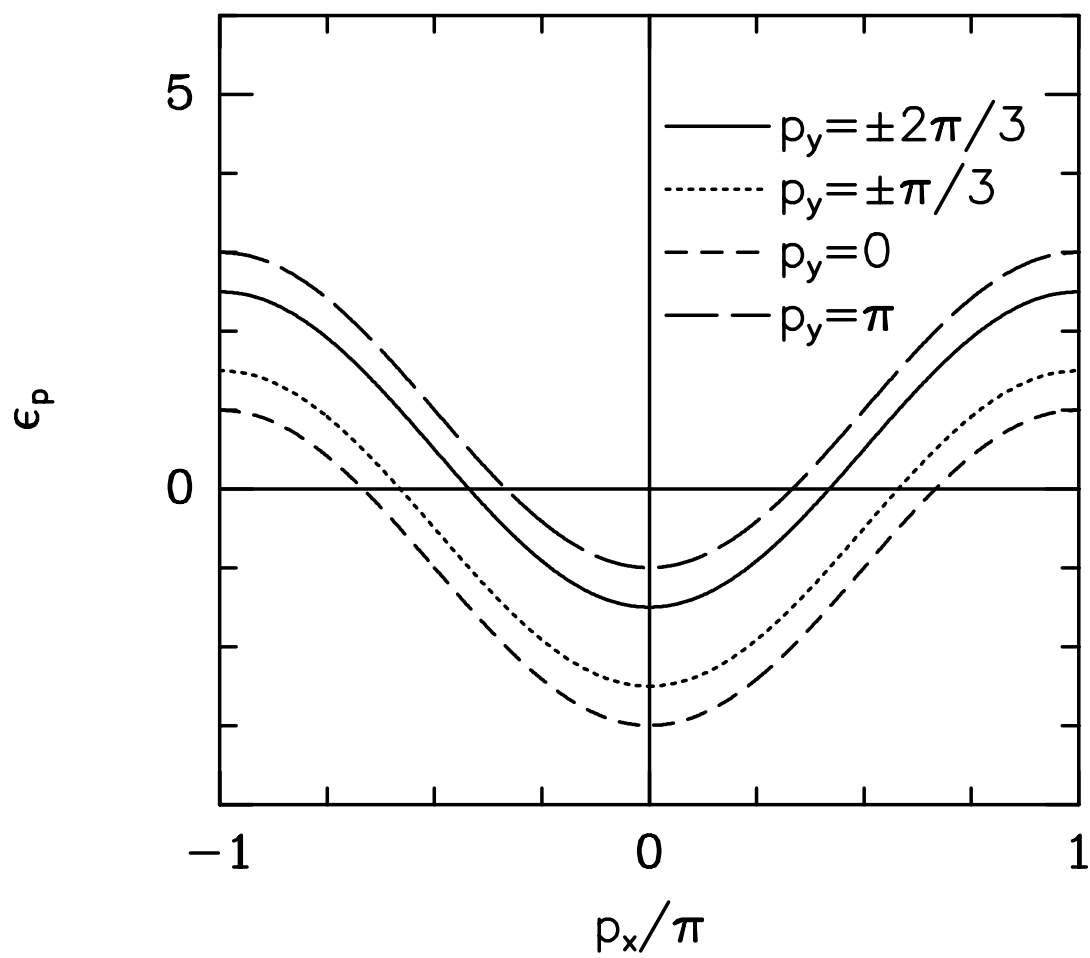
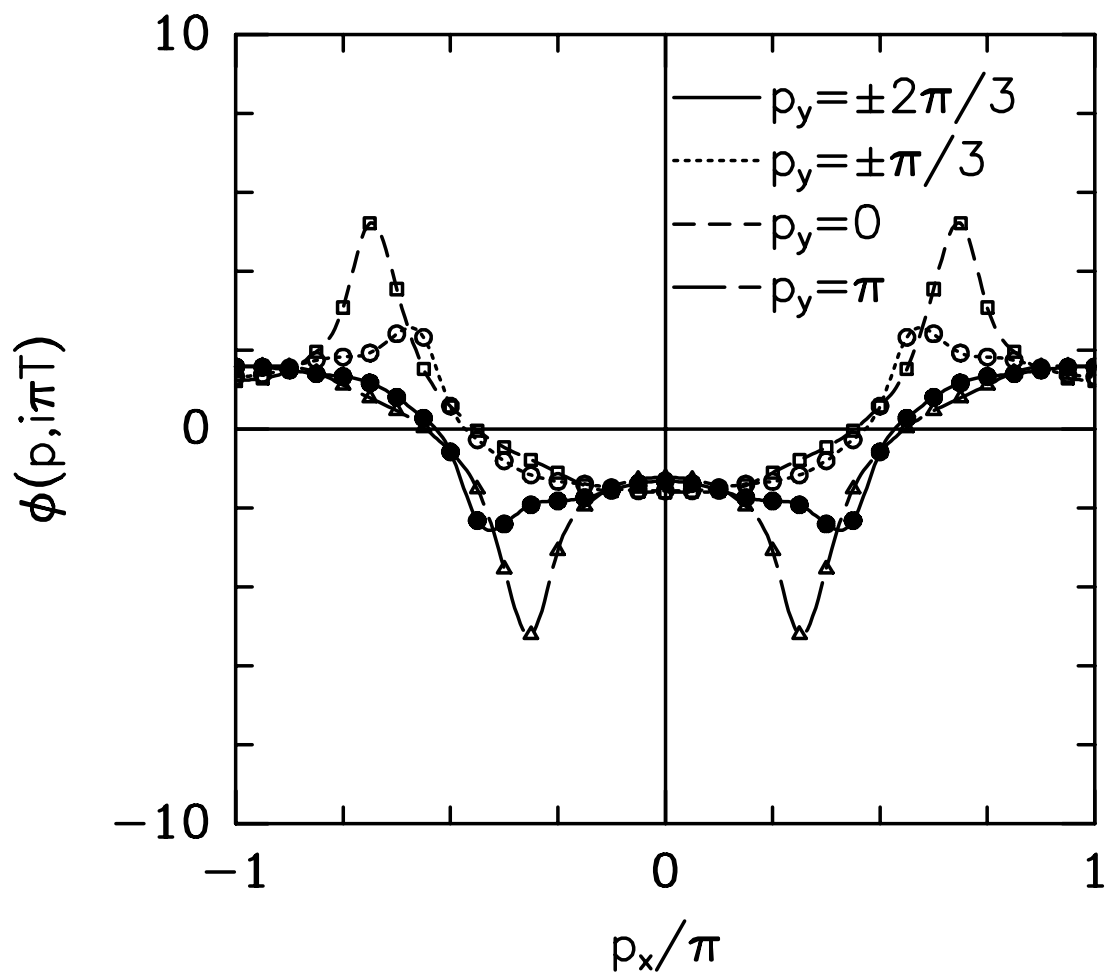
\begin{figure}
\caption{
(a) Quasiparticle dispersion $\varepsilon_{{\bf p}}$, and
(b) $\phi({\bf p},i\pi T)$ versus $p_x$
for different values of $p_y$.
These results are for a six--leg ladder
with $U=1.5t$, $t_{\perp}=0.5t$,
$\langle n\rangle = 1.0$, and $T=0.1t$.
}
\label{fig7}
\end{figure}
  
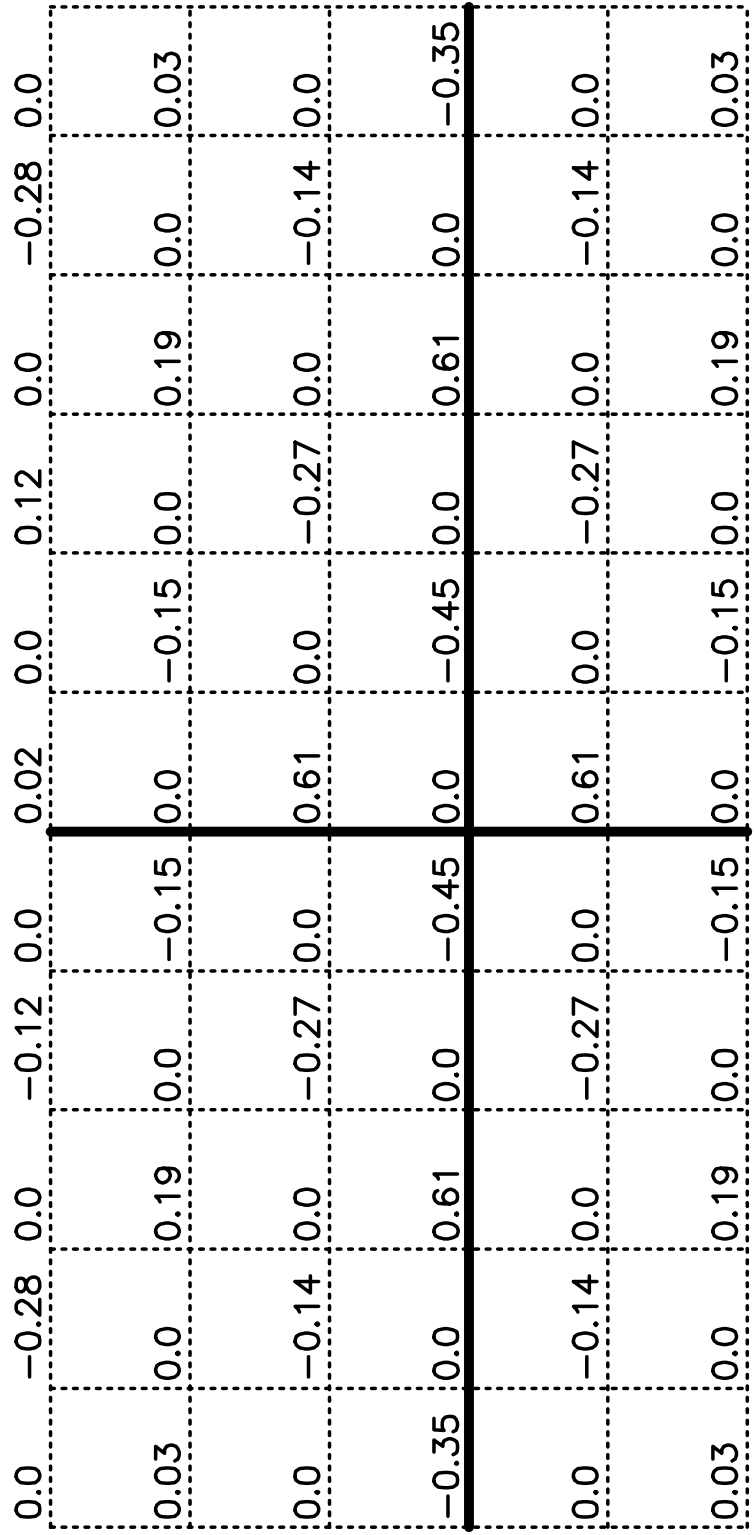
\begin{figure}
\caption{
Pair amplitude $A(\ell_x,\ell_y)$ versus $(\ell_x,\ell_y)$
for the same parameters as in Fig. 7.
}
\label{fig8}
\end{figure}
  
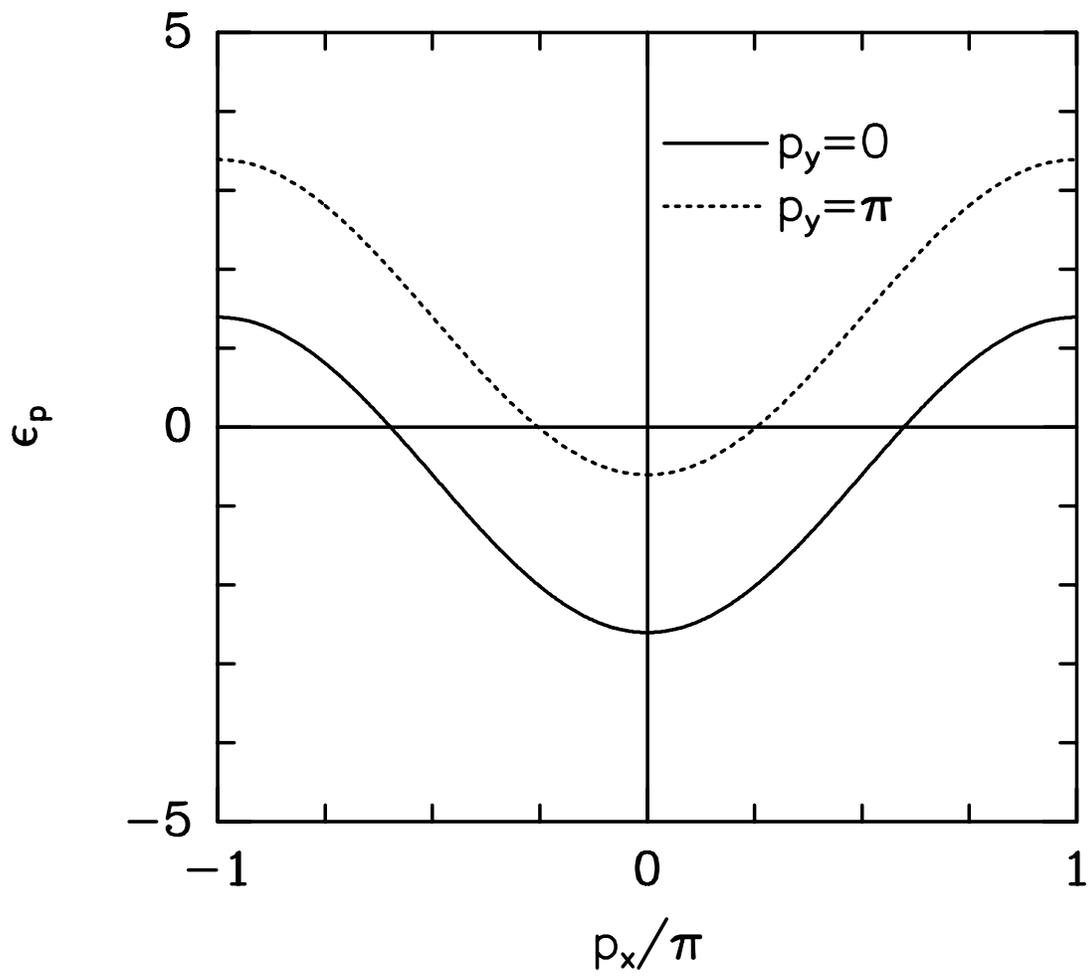
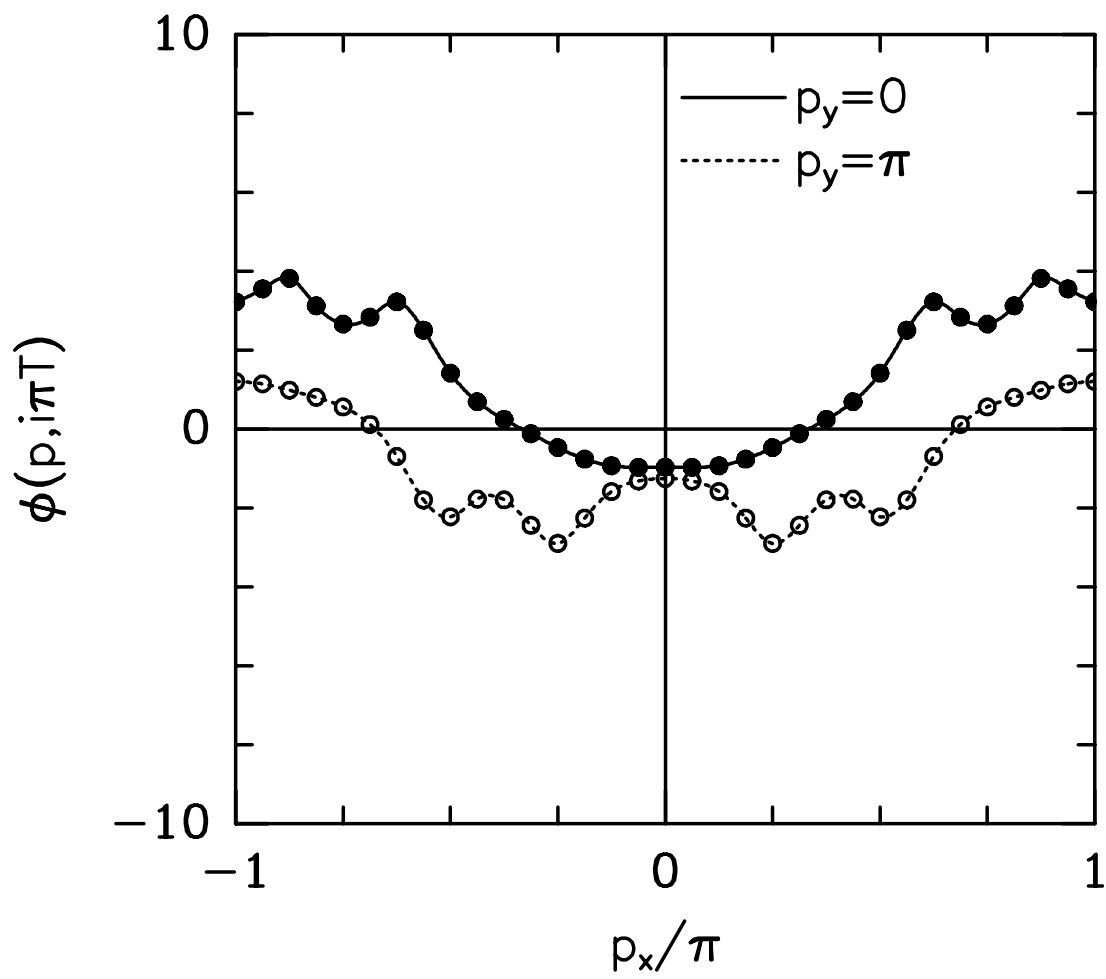
\begin{figure}
\caption{
(a) Quasiparticle dispersion $\varepsilon_{{\bf p}}$, and
(b) $\phi({\bf p},i\pi T)$ versus $p_x$
for $p_y=0$ and $\pi$.
These results are for a two--leg ladder 
with $U=2.0t$, $t_{\perp}=0.5t$,
$\langle n\rangle = 0.85$, and $T=0.1t$.
}
\label{fig9}
\end{figure}
  
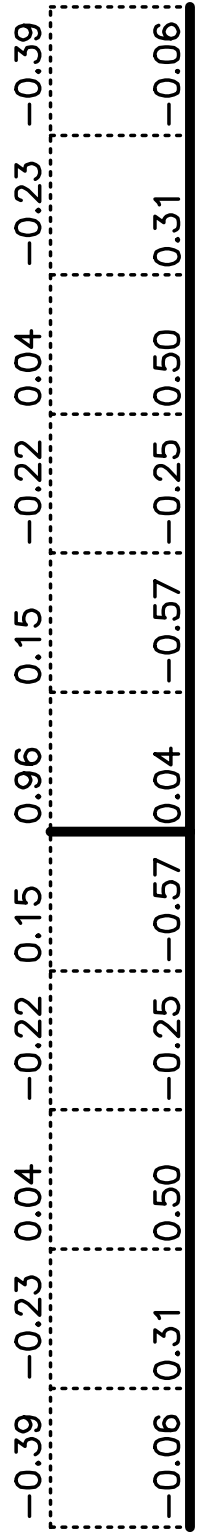
\begin{figure}
\caption{
Pair amplitude $A(\ell_x,\ell_y)$ versus $(\ell_x,\ell_y)$
for the same parameters as in Fig. 9.
}
\label{fig10}
\end{figure}
  
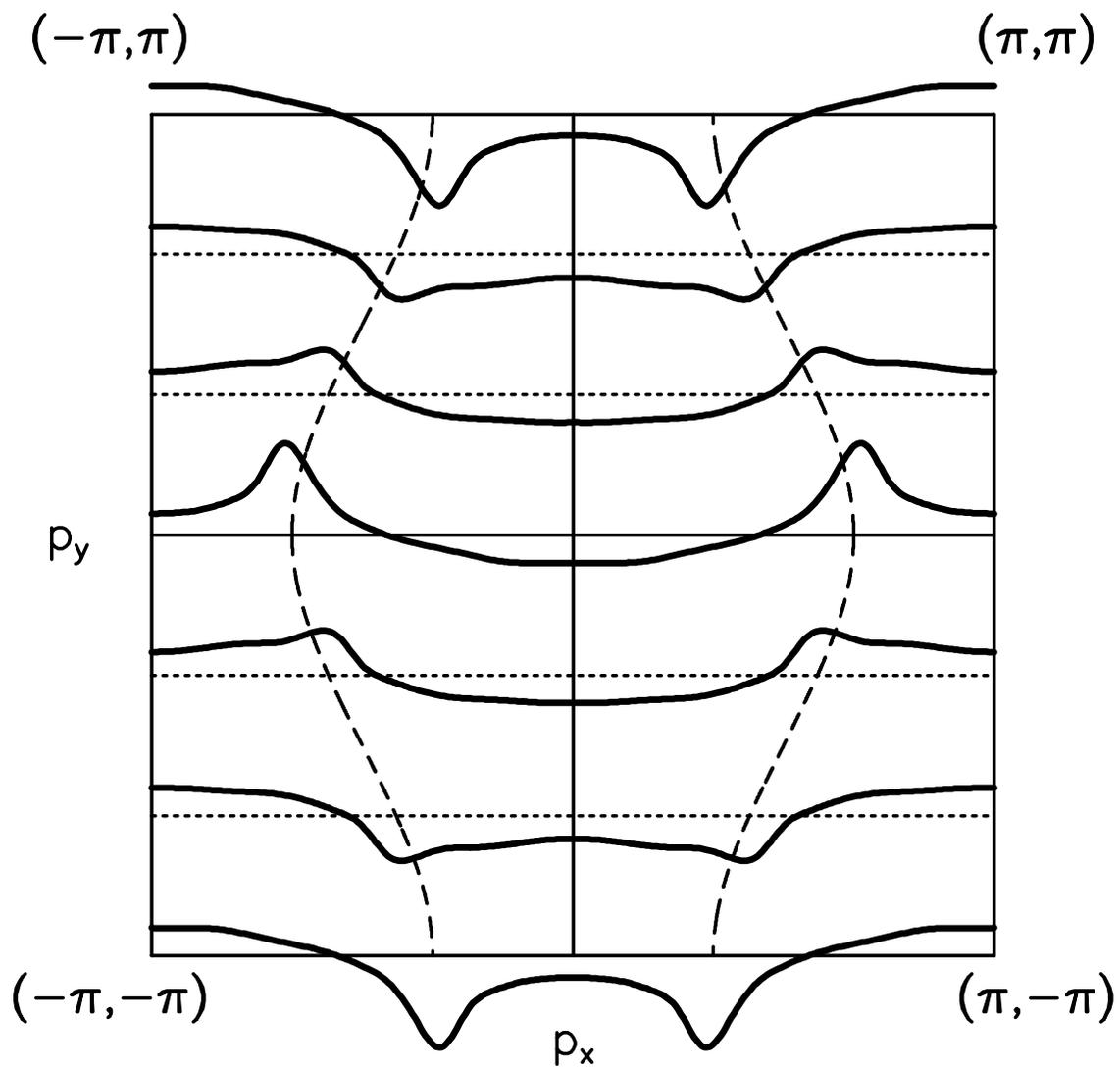
\begin{figure}
\caption{
$\phi({\bf p},i\pi T)$ (thick solid curves) 
versus $p_x$ for $p_y=0$, $\pm \pi/3$, $\pm 2\pi/3$, and $\pm \pi$.
These results are for a six--leg ladder
with $U=1.5t$, $t_{\perp}=0.5t$, $T=0.1t$,
and $\langle n\rangle = 1.0$.
Here,
the dashed curves indicate the fermi surface of the 
two dimensional lattice with $t_{\perp}=0.5t$.
}
\label{fig11}
\end{figure}

\end{document}